\documentclass[pra,amssymb,amsfonts,12pt]{revtex4}
\usepackage{bm,bbm}
\usepackage{graphicx}

\begin{document}

\title{Basis Markov  Partitions and Transition Matrices \\ 
   for Stochastic Systems}
\author{Erik Bollt$^{1}$, Pawe{\l} G{\'o}ra$^{2}$,
  Andrzej Ostruszka$^{3}$, and Karol \.Zyczkowski$^{3,4}$
\medskip \\
$^1${\small Department of Mathematics \& Computer Science and Department of
Physics, Clarkson University, Potsdam, NY 13699-5805 } \\
$^2${\small Department of Mathematics and Statistics, Concordia University 
7141 Sherbrooke Street West, Montreal, Quebec H4B 1R6, Canada} \\
$^3${\small Institute of Physics, Jagiellonian University, 
           ul. Reymonta 4, 30--059 Krak\'ow, Poland} \\
$^4${\small Center for Theoretical Physics, Polish Academy of Sciences, \\
 Al. Lotnik\'ow 32/44, 02--668 Warszawa, Poland} \\
 }

\email{bolltem@clarkson.edu, 
         pgora@vax2.concordia.ca, 
         karol@tatry.if.uj.edu.pl}


\date{May 8, 2006}

\begin{abstract}
{We analyze dynamical systems subjected to an additive noise and their 
 deterministic limit. In this work, we will introduce a notion by which a stochastic system has
something like a Markov partition for deterministic systems. For a chosen class of the noise profiles
 the Frobenius-Perron operator associated to the noisy system 
 is exactly represented by a stochastic transition matrix of a finite size $K$. 
 This feature allows us to introduce for these stochastic systems
 a basis--Markov partition, defined herein, irrespectively of whether the deterministic 
 system possesses a Markov partition or not. We show that in the deterministic  
 limit, corresponding to $K \to \infty$, the sequence of invariant measures
 of the noisy systems tends, in the weak sense, to the invariant measure
 of the deterministic system. Thus by introducing a small
 additive noise one may approximate transition matrices and 
 invariant measures of deterministic dynamical systems. 
}
\end{abstract}
\maketitle

\section{Introduction}

 Markov partitions for deterministic dynamical systems serve a central role for
determining their symbolic dynamics, \cite{Bowen75,BG,bollt33} whose
grammar is described  by a finite sized transition matrix that generates a
so-called sofic shift \cite{Kitchens, Fischer}. 
The conditions for such a
projection were defined by Bowen for Anosov hyperbolic systems
\cite{Bowen75,bollt33}, and stated succinctly for interval maps as a partition
whose elements are each a homeomorphism onto a finite union of its elements
\cite{BG,bollt33}.  
We remark here that a defining property in both cases is
that the set of characteristic functions defined over the elements of the
Markov partition project the transfer operator  exactly onto an operator of
finite type; that is a matrix results whereas an infinite matrix would be
expected for a non-Markov system.  We argue here that this should  be the
defining property of any generalization of Markov partitions; that is, a set of basis functions
which project the Frobenius-Perron operator exactly onto a finite-rank matrix
with no residual.

First we recall the Frobenius-Perron operator for a deterministic
transformation. Associated with a discrete dynamical system acting on initial
conditions, $\mathbf{z} \in M$ (say a manifold, $M\subset \Re^n$), 
\begin{eqnarray}
F:M &\rightarrow& M, \nonumber \\
\mbox{ } x &\mapsto&  F(x),
\end{eqnarray}
is another dynamical system over $L^1(M)$, the space of densities of
ensembles of initial conditions.
\begin{eqnarray}
\label{FPD}
P_F:L^1(M) &\rightarrow&  L^1(M),\nonumber \\
 \mbox{ }\rho(x) &\mapsto&
P_F[\rho(x)]. 
\end{eqnarray}
This Frobenius-Perron operator $(P_F)$ is defined through a continuity equation
\cite{Lasota-Mackey},
\begin{eqnarray}
\int_{F^{-1}(B)} \rho(x) dx=\int_B P_F[\rho(x)] dx,
\end{eqnarray}
for measurable sets $B\subset M$. Differentiation changes this operator
equation to the commonly used form, 
\begin{equation}
\label{FPD2}
P_F[\rho(x)]=\int_M \delta(x-F(y)) \rho(y) dy,
\end{equation}
acting on probability density functions $\rho\in L^1(M)$.

Now consider the stochastically perturbed dynamical system
\begin{eqnarray}\label{smap}
F_\nu:M & \rightarrow&  M, \nonumber \\
 \mbox{ } x &\mapsto & F(x)+\xi,  
\end{eqnarray}
where $\xi$ is an i.i.d. random variable with PDF $\nu(z)$, which is applied
once per each iteration. The random part $\nu$ is assumed to be independent of
state $x$ which we tacitly assume to be relatively small, so that the
deterministic part $F$ has primary influence.
The ``stochastic Frobenius-Perron operator'' has a similar form to the
deterministic case \cite{Lasota-Mackey},
\begin{equation}
\label{FPnoise2}
P_{F_\nu}[\rho(x)]=\int_M \nu(x-F(y)) \rho(y) dy,
\end{equation}
where the deterministic kernel, the delta function in Eq.~(\ref{FPD}), now
becomes a stochastic kernel describing the PDF of the noise perturbation.
In the case that the random map Eq.~(\ref{smap}) arises from the usual
continuous Langevin process, the infinitesimal generator of the
Frobenius-Perron operator (FP--operator) for normal $\nu$ corresponds to a
general solution
of a Fokker-Planck equation, \cite{Lasota-Mackey}.  The Frobenius-Perron
operator formalism is particularly convenient in that it allows for an
arbitrary noise distribution $\nu$  to be incorporated in a  direct and simple
way.   Within the formalism, we can also study multiplicative noise
($x\rightarrow \eta F(x)$, modeling parametric noise).   The kernel-type
integral transfer operator is, ${\cal K}(x,y)=\nu(x/F(y))/F(y)$ for $x\in
\Re^+$,
which can then also be finitely approximated as described in the next section,
and usefully  re-ordered to canonical block reduced form.  In more generality,
the theory of random dynamical systems \cite{LArnold} clearly classifies those
random systems which give rise to explicit transfer operators with
corresponding infinitesimal generators, and there are well defined connections
between the theories of random dynamical systems and of stochastic
differential equations.  

The main aim of this work is to investigate a class of 
stochastically perturbed dynamical systems for which
the FP operator is represented by a finite stochastic
transition matrix of size $K$.
Such dynamical systems will be called {\sl basis--Markov}
in analogy to deterministic
dynamical systems possesing a Markov partition,
for which the associated FP operator is finite.
The deterministic limit of the stochastic system
corresponds to the divergence of the matrix size.
In this limit, $K\to \infty$,
the sequence of invariant measures
of the stochastic systems acting in the $K$--dimensional Hilbert space
converges, in the weak sense, to the invariant measure 
of the coresponding deterministic system.

The paper is organized as follows. 
The Ulam-Galerkin method of approximating the infinite dimensional
FP operator and the concept of the Markov partition for a
deterministic system are reviewed in sections II  and III, respectively.
In section IV we introduce the notion
of basis-Markov stochastic systems,
while in section V we analyze a particular
example of random systems perturbed by an additive noise 
with cosine profile.
The key result on convergence of the invariant measures
for stochastic and deterministic systems is proved in section VI.
A discussion
of isospectral matrices used to describe the FP operator
is relegated to the appendix.

\section{Ulam-Galerkin's Method-Approximating the Infinite-Dimensional
Operator}

A Galerkin's method may be used to approximate the Frobenius-Perron operator by
a Markov operator of finite rank. Formally, projection of the infinite
dimensional linear space $L^1(M)$ results with discretely indexed basis
functions $\{\phi_i(x)\}_{i=1}^\infty \subset L^1(M)$ onto a finite
dimensional linear subspace generated by a subset of the basis functions
\cite{Li}, 
\begin{equation}
\Delta_N =span(\{\phi_i(x)\}_{i=1}^N),
\end{equation}
 such that $\phi_i\in L^1(M)$ $\forall i$. This projection, 
 \begin{equation}
 p:L^1(M)\rightarrow \Delta_N,
 \end{equation}
  is realized optimally by the Galerkin method in terms of the inner product,
which we choose to be integration, 
  \begin{equation}
  (f,g)\equiv \int_M f(x) g(x) dx, \forall f,g\in L^2(M).
  \end{equation}
   Specifically, the infinite-dimensional ``matrix" is approximated by the
$N\times N$ matrix,
\begin{equation}\label{Gally}
A_{i,j}=(P_{F_\nu}[\phi_i],\phi_j)=\int_M
P_{F_\nu}[\phi_i(x)]\phi_j(x) dx, \mbox{ } 1\leq i,j\leq N.
\end{equation}
One approximates $\rho(x)$, through a finite linear combination of basis
functions, 
\begin{equation}
\rho(x)\simeq \sum_{i=1}^N d_i \phi_i(x).
\end{equation}
The historically famous Ulam's method \cite{Ulam} for deterministic dynamical
systems is equivalent to the interpretation to find  the fraction of the box
$B_i$ which maps to $B_j$; the Ulam matrix is  equivalent to the Galerkin
matrix by using Eq.~(\ref{Gally}) and choosing the basis functions to be the
family of characteristic functions,
\begin{equation}
\phi_i(x)={\mathbf 1}_{B_i}(x)=
\left(\begin{array}{ccc} 1 &\mbox{ if } &x\in B_i \\ 0 & \mbox{ } & \mbox{
else}.\end{array}\right)
\end{equation}
Specifically, we choose the ordered set of basis functions to be in terms of a
nested refinement of boxes $\{B_i\}$ covering $M$. Though Galerkin's and
Ulam's methods are formally equivalent in the deterministic case, we are of
the opinion that the Galerkin description is a more natural description in the stochastic
setting.

\section{Markov Partitions of Deterministic Systems, and Exact Projection}

In this section, we discuss that a Markov partition is special for the
Frobenius-Perron operator of a deterministic dynamical system, in that
characteristic functions supported over those partition elements leads to an
exact projection of the FP operator onto an operator of finite rank - a
matrix.

For a one-dimensional transformation of the interval, a Markov partition is
defined \cite{BG, bollt33},

\medskip \noindent
{\bf Definition:} A map of the interval $f:[a,b] \rightarrow [a,b]$ is Markov
if there is a finite partition $\{I_j\}$ such that,
\begin{enumerate}
\item $\cup_j I_j =[a,b]$ (covering property),
\item ${\sl int}(I_j)\cap {\sl int}(I_k)=\emptyset \mbox{ if } k\neq k$ (no overlap
property),
\item $f(I_j)=\cup_{k_i} I_{k_i},$ (a grid interval maps completely across a
union of intervals without ``dangling ends" property).
\end{enumerate}

It is not hard to show that the set of characteristic functions forms a finite
basis set of functions
\begin{equation}
\{\phi_i(x)\}=\{{\mathbf 1}_{I_i}(x)\}_i,
\end{equation}
 such that Galerkin projection Eq.~(\ref{Gally}) is exact onto an operator of
finite rank, or a matrix $A_{i,j}$.  That is, Eq.~(\ref{Gally}) simplifies, 
\begin{eqnarray}\label{finiterank}
A_{i,j}&=&(P_{F_\nu}[\phi_i],\phi_j)=\int_M
P_{F_\nu}[\phi_i(x)]\phi_j(x) dx, \nonumber \\
&=& 
\int_M \int_M \delta(x-F(y)) \phi_i(y) \phi_j(x) dy dx \nonumber \\
&=&
 \int_{I_j} \int_{I_i} \delta(x-F(y))  dy dx  \mbox{ } 1\leq i,j\leq N.
\end{eqnarray}
From the definition of the Markov partition, we see that a
 row of $A_{i,j}$ accounts that $P_{F_\nu}[\phi_i(x)]$ is a linear combination
of $\phi_{j}(x)$.

 Similarly, there is a well defined notion of an Anosov diffeomorphisms with a
Markov partition \cite{Bowen75,bollt33,Robinson,GuckHolmes}, and so for such
systems, it can be shown that characteristic 
functions supported over the corresponding Markov partition creates a basis set
such that Eq.~(\ref{Gally})  results in an operator of finite rank.

We  take these observations as motivation to make the following definition
which is meant to generalize the notion of a Markov partition to stochastic
systems:

\medskip
\noindent
{\bf Definition:} 
Suppose a measure space, $\{M,{\cal B},\mu\}$, and a transformation
$F:M\rightarrow M$, 
then the transformation is ``{\bf basis  Markov}" if there exists a finite set
of 
basis functions $\{\phi_i(x)\}_{i=1}^n:M\rightarrow [0,1]\in L^1(M)$ such that
the Frobenius-Perron 
operator is operationally closed within $\Delta_n$, where
$\Delta_n=span(\{\phi_i(x)\}_{i=1}^n)$.  
That is, 
for any probability measure $\rho$, its image 
$P_F[\rho(x)]$ belongs to $\Delta_n$.

\medskip
\noindent
{\bf Remark 1:}   If a transformation $F$ is basis-set  Markov, then if we
perform Galerkin's method, $A_{i,j}=(P_{F_\nu}[\phi_i],\phi_j)_{M\times M}$,
with that basis set, then it allows that for any initial density which can be
written as a linear combination of these basis functions, 
\begin{equation}
\rho_0(x)=\sum_{i=1}^n c_i \phi_i(x),
\end{equation}
or stated simply,
\begin{equation}
\rho_0(x)\in \Delta_n,
\end{equation}
 then the action of the Frobenius-Perron operator on such initial densities, 
$\rho_1(x)=P_{F_\nu}[\rho_0(x)]$, can be exactly represented by the following
matrix-vector multiplication: 
\begin{equation}\label{matrixapprox}
{\mathbf c}'=A\cdot {\mathbf c}, \mbox{ where }\rho_1(x)=\sum_{i=1}^n c_i'
\phi_i(x).
\end{equation}
  That is, the FP operator projects exactly to an operator of finite rank - a
matrix.
\medskip

Note that for a general finite set of functions, if we take a general linear
combination of those functions and then apply the Frobenius-Perron operator,
we do not expect the resulting density can be written as a (finite) linear
combination of basis functions.  

\medskip
\noindent

The following is a direct consequence of our definition of basis Markov in
relationship to the usual definition of a Markov map, stating the sense in
which basis Markov is a generalization:

\medskip
\noindent
{\bf Remark 2:} Given a Markov map, then Eq.~(\ref{finiterank}) implies that
any Markov map, together with the characteristic functions supported over the
partition elements, is basis Markov.

\section{Basis Markov stochastic systems: A General Case Due to Separable
Noise}

We analyze a dynamical system defined on an interval $M=[0,1]$ 
with both ends identified and
subjected to a specific form of the additive noise,
\begin{eqnarray}
 x'=f(x) + \xi \: .
\label{addit}
\end{eqnarray}
To specify the special case of the stochastic dynamical system written in
Eq.~(\ref{smap}),
the stochastic perturbation will be characterized 
by the probability ${\cal P}(x,y)$
of a transition form point $x$ to $y$ induced by noise.
Describing the dynamics in terms of a probability density $\rho(x)$
its one-step evolution is governed 
by the stochastic {\sl Frobenius-Perron (FP) operator},
\begin{equation}\label{prob1}
\rho'(y) = P_f\bigl( \rho(y)\bigr) \:  = \:  \int{\cal P}\bigl( f(x),y\bigr) \, 
\rho(x)dx.
\label{FPnoise}
\end{equation}
We will denote this stochastic Frobenius-Perron operator by the symbol $P_f$,
in all that follows.
The operator $P_f$ acts on every probability measure defined on $M$
and in general, it cannot be represented by a finite matrix.
However, in the sequel we shall analyze a certain class of noise profiles
for which such a representation is possible.

 We assume that the transition probability ${\cal P}(x,y)$
satisfies the following properties \cite{OPSZ00,OZ01}:
\begin{eqnarray}\label{prob2}
{\rm a)} \quad {\cal P}(x,y) & \equiv & {\cal P}(x-y)={\cal P}(\xi),  \nonumber
\\
{\rm b)} \quad {\cal P}(x,y) & \equiv & {\cal P}(x\bmod1,y\bmod1), \nonumber \\
{\rm c)} \quad {\cal P}(x,y) & = &  \sum_{m,n=0}^{N}\, A_{mn}\, u_n(x)\,
v_m(y),
\label{expand}
\end{eqnarray}
for $x,y \in {\mathbb R}$ and an arbitrary finite $N$.
Property a) assures that  
the distribution of the random variable $\xi$ 
does not depend on the position $x$, 
while the periodicity condition is provided in b).
A noise profile fulfilling the latter property c)
is called {\sl separable} (decomposable),
and it allows us to represent the dynamics
of an arbitrary the system with such a noise 
in a finite dimensional Hilbert space. 
Here $A = (A_{mn})_{m,n = 0, \dots ,N}$ is a yet undetermined 
real matrix of expansion coefficients. 
Note that $A$ characterizes the noise and 
does not depend on the deterministic dynamics $f$.
We assume that the functions $u_n; ~n = 0,\dots,N$ and $v_m; ~m = 0,\dots,N $
 are continuous in $X=[0,1)$ and linearly independent, 
so we can express $f \equiv 1$ as their linear combinations.
Both sets of functions span bases in an $N+1$ Hilbert space. 
Their orthogonality is not required.

This name ``separable noise'' is concocted in an analogy 
to {\sl separable states} in quantum mechanics 
and {\sl separable} probability distributions,
since such a property was called $N+1$-separability 
by Tucci \cite{Tu00}.
Making use of this crucial feature of the noise profile
we may expand the 
kernel of the Frobenius--Perron operator  (\ref{FPnoise}),
\begin{eqnarray}
\rho'(y)=
P_f\bigl( \rho(y)\bigr) & = & 
 \int_0^1  \sum_{m,n=0}^N A_{mn} u_n(f(x)) v_m(y) \rho(x) dx
\\
& =  & \sum_{m,n=0}^N A_{mn} 
\Bigl[ \int_0^1 u_n(f(x)) \rho(x) dx \Bigr] v_m(y) 
\\
&=& \sum_{n=0}^N \Bigl[ \int_0^1 u_n(f(x)) \rho(x) dx \Bigr]
 \tilde{v}_{n}(y) \nonumber 
\label{fp1}
\end{eqnarray}
for $y \in X$, where,
\begin{equation}
\tilde{v}_n=\sum_{m=0}^NA_{mn}v_m.
\end{equation}
 Thus, any initial density is projected by the FP--operator $P_f$ into 
the vector space spanned by the functions $\tilde{v}_{m}; m=0,\dots,N$.

Assuming that a given density $\rho(x)$ belongs to this space, 
we can be expand it in this basis,
\begin{equation}
\rho(x) = \sum_{m=0}^N {q}_m\, \tilde{v}_m(x) \ .
\label{expmes}
\end{equation}
Expanding $\rho'$ in an analogous
way we will describe it by the vector ${\vec q}'=\{q_0',\dots,q_N'\}$. 

Let $B$ denotes a matrix of integrals,  
\begin{equation}
B_{nm} = \int_0^1 u_n(f(x)) v_m(x) dx ,
\label{brm}
\end{equation}
where  $n,m=0,\dots,N$. 
Observe that  $B$ depends directly on the system $f$
and on the noise via the basis functions $u$ and $v$.
Making use of this matrix, 
the one--step dynamics (\ref{fp1}) 
may be rewritten in a matrix form  
\begin{equation}
q'_n  = \sum_{m=0}^N \, D_{nm}\, q_m ,
{\quad \rm where \quad} 
D=BA\,
\label{fp2}
\end{equation}
and  $A$ is implied by (\ref{expand}).
In this way we have arrived at 
a representation of the Frobenious--Perron
operator $P_f$ by a matrix $D$ of size $N+1 \times N+1$,
the elements of which read,
\begin{equation}
D_{nm} = \int_0^1 u_n(f(x)) \tilde{v}_m(x) dx, \quad
n,m=0,\dots,N.
\label{drm}
\end{equation}

Although the probability is conserved
under the action of $P_f$, the matrix $D$ need not 
be stochastic. This is due to the fact that 
the  functions $\{  \tilde{v}_m (x)\}$ 
forming the expansion
basis in (\ref{expmes}) were not normalized.
We shall then compute their norms, 
\begin{equation}
\tau_m= \int_0^1 \tilde{v}_m(y)dy = \sum_{n=0}^N A_{mn}b_n
\label{norm1}
\end{equation}
where,
\begin{equation}
b_n= \int_0^1 v_n(y)dy.
\end{equation}
 Let  $K\le N+1$ denote the number of non-zero components 
of the vector  ${\vec \tau}$
and  let $k=1,\dots,K$ runs over all indexes $n\in 0,\dots N+1$,
for which $\tau_k\ne 0$.
Then the rescaled vectors,
\begin{equation}
V_k(y):=\tilde{v}_k(y)/\tau_k,
\end{equation}
are normalized, 
\begin{equation}
\int_0^1 V_k(y)dy=1.
\end{equation}

The normalization condition $\int_0^1 \rho(x)dx =1$ implies
\begin{equation}
  \int_0^1 \sum_{m=0}^N q_l \tilde{v}_m(x)dx = \sum_{m=0}^N q_m \tau_m
= \sum_{k=1}^K q_k \tau_k =1
\label{norm11}
\end{equation}
The same is true for the transformed density,
\begin{equation}
\sum_k q_k' \tau_k =1.
\end{equation}
Hence this scalar product is preserved
during the time evolution.  Making use of the rescaled coefficients 
\begin{equation}
c_k:=q_k \tau_k,
\end{equation}
the  dynamics (\ref{fp2})
reads
\begin{equation}
c_k'=q_k' \tau_k =\sum_j D_{kj} \, q_j \tau_k =
\sum_j D_{kj} \, \frac{\tau_k}{\tau_j} \, q_j \tau_j =: 
\sum_j T_{kj}\, c_j .
\label{norm2}
\end{equation}
The coefficients $c_k$ sum to unity, so the
transition matrix
\begin{equation}
T_{kj} \ \equiv \ D_{kj} \, \frac{\tau_k}{\tau_j} =
\sum_{ii'}
 D_{kj} \, \frac{A_{ki} \tau_i}{A_{ji'} \tau_{i'}} \ .
\label{norm22}
\end{equation}
is stochastic.
In the above equation, all indices run from $1$ to $K$
and the coefficients $\tau_k$ are non-zero by construction.
Hence the dynamics
(\ref{fp2}) effectively takes place in an $K$-dimensional Hilbert space,
and the Frobenious--Perron operator $P_f$ is represented
by a stochastic matrix $T$ is of size $K\times K$.
The dimensionality $K\le N+1$ is determined by the parameter $N$
and the choice of the basis functions $\{v_l(x)\}$
 entering  (\ref{expand}).

\section{A special case: cosine noise}

\begin{figure}
\includegraphics[height=3.in]{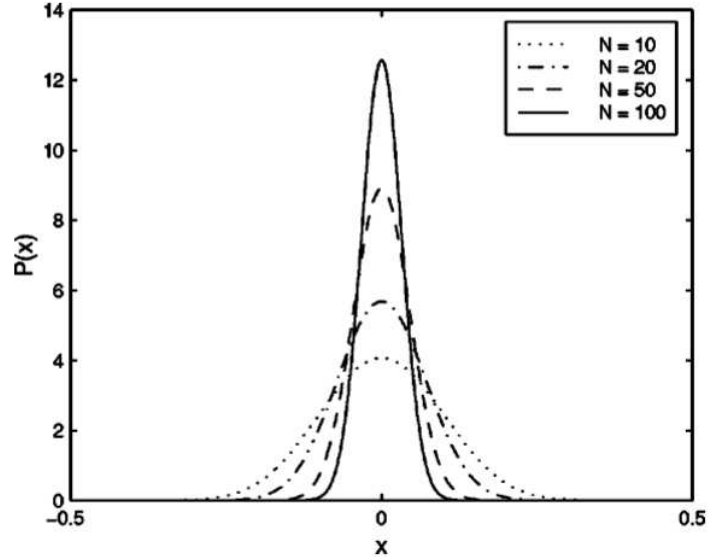}
\caption{The cosine noise of Eq.~(\ref{trignoise}) closely resembles a normal
noise profile, but with finite support.  Several values of $N$ are shown, with
decreasing standard deviation with increasing $N$.  
} \label{figure0}
\end{figure}

We will now discuss a particularly simple case of the 
separable noise described above, introduced in \cite{OPSZ00}.  Let,
\begin{equation}
\label{trignoise}
{\cal P}_N(\xi) = {\rm C_N}\cos^N(\pi\xi)\,,
\end{equation}
where $N$ is even ($N=0,2,\dots$) and 
with the normalization constant,
\begin{equation}
{\rm C_N} = \sqrt{\pi}  \, \frac{ \Gamma[N/2+1] }
                         {\Gamma[(N+1)/2] } \ . 
\label{cnorm}
\end{equation}
See Fig.~\ref{figure0}, in which we can see the decreasing standard deviation with
respect to increasing $N$, and it can be seen that this type of noise reminds
of a normal distribution, but of compact support.

The parameter $N$ controls the strength of the noise measured by its variance
\begin{equation}
\sigma^2 = \frac{1}{2\pi^2}\Psi'(\frac{N}{2}+1)
 = \frac{1}{12} - \frac{1}{2 \pi^2}
 \sum\limits_{m=1}^{N/1} \frac{1}{m^2} \ ,
\end{equation}
where $\Psi'$ stands for the derivative of the digamma function.

For the expansion  (\ref{expand})
 we use basis functions,
\begin{eqnarray}\label{trigbasis}
u_m(x) &=& \cos^m(\pi x) \sin^{N-m}(\pi x), \nonumber\\
v_n(y) &=& \cos^n(\pi y) \sin^{N-n}(\pi y),
\end{eqnarray}
where $x \in X$ and $m,n = 0,\dots,N$.
 Expanding cosine as a sum to the $N$-th power in
   (\ref{trignoise}) we find that the $(N+1) \times (N+1)$
    matrix $A$ defined by (\ref{expand})   is diagonal, 
\begin{equation}
A_{mn}= a_m  \delta_{mn}
\mbox{, with }
a_m = {\rm C_N} {N \choose m} \ .
\label{alr}
\end{equation}

\begin{figure}
\includegraphics[height=4.in]{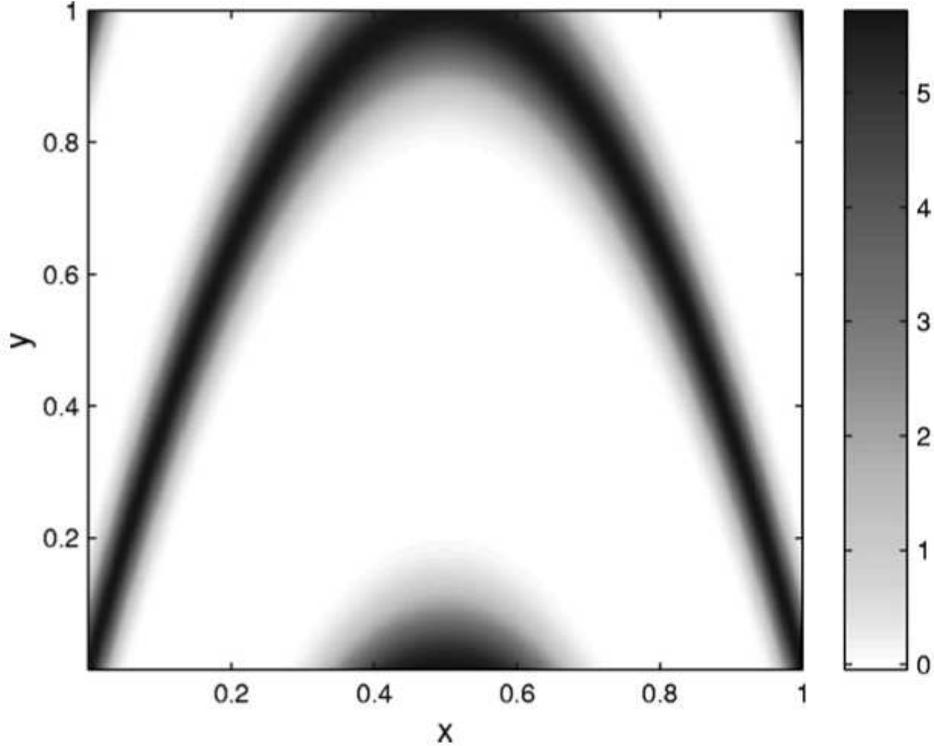}
\caption{ 
The transition kernel ${\cal P}_N(f(x),y)$ for the logistic map $f(x)=4x(1-x)$,
with $N=20$
  and with cosine noise due to $N=20$; compare to Fig.~\ref{figure0}. Note the
periodicity of $x$ of period-1.
 }
\label{figure1}
\end{figure}

\begin{figure}
\includegraphics[height=4.in]{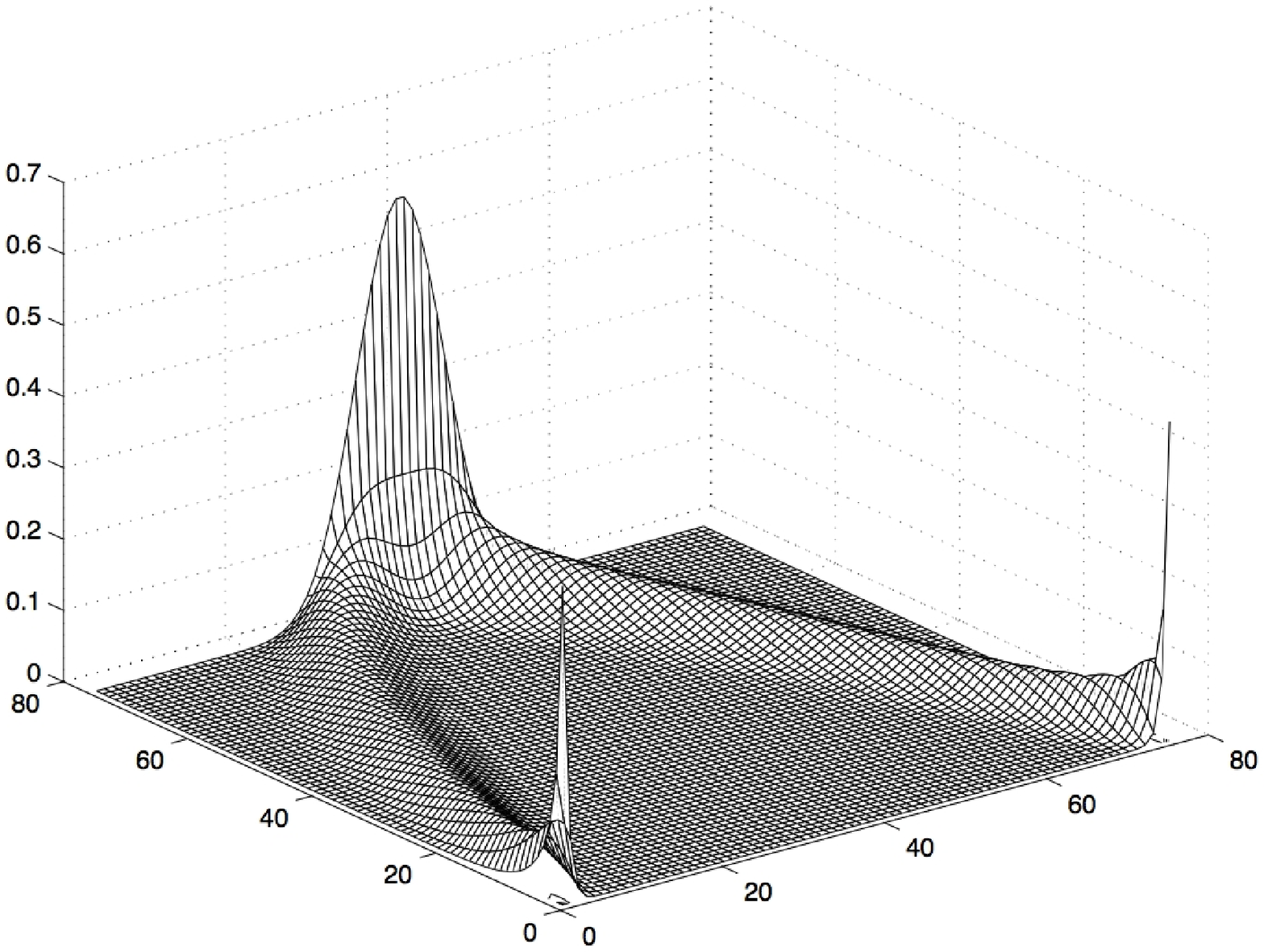}
\caption{The stochastic matrix $T_{150}$ shown, from Eq.~(\ref{norm22}),
exactly represents the stochastic Frobenius-Perron operator of the stochastic
tent map Eq.~(\ref{tent}) with trig noise Eq.~(\ref{trignoise}), and basis set
Eq.~(\ref{trigbasis}), using $N=150$.  Note that $T^{(150)}$ is a matrix of
size $N/2+1=150/2+1=76$ square. Compare to the matrices in Eq.~(\ref{TTT2}),
of smaller $N$.  
} \label{figure2}
\end{figure}

Integrating  trigonometric functions we find the 
coefficients,  
\begin{equation}
b_{m} = \int_0^1 \sin^m (\pi x) \cos^{N-m} (\pi x) dx = 
\frac{2}{\pi N}  \,
\frac{\Gamma[(m+1)/2]\,  \Gamma[(N-m+1)/2] }
       {\Gamma(N/2)} \ ,  
\label{coebn}
\end{equation}
and,
\begin{equation}
\tau_m=a_m b_m,
\end{equation}
which are non-zero only for even values of $m$.
Hence the size $K\times K$ of the transition matrix 
reads,
\begin{equation}
K=N/2+1,
\end{equation}
 and 
the expression (\ref{norm22}) takes the form 
\begin{equation}
T_{kj}  =  D_{mn} \, \frac{a_m b_m }{a_n b_n} 
{\rm \quad where \quad} k,j=1,\dots, K; \ \ 
m=2(k-1), \ \ n=2(j-1) \  .
\label{tstoch}
\end{equation}
We find in the cosine noise Eq.~(\ref{trignoise}) and with basis
Eqs.~(\ref{trigbasis}), that the transition kernel reminds of a fuzzy but
periodically repeated version of the map.  See Fig.~\ref{figure1}.  However,
the Frobenius-Perron operator embeds to a transition matrix $T$, which
``appears" roughly as a different form of the original map.  See
Fig.~\ref{figure2}.  However, with zero-noise, 
an Ulam transition matrix approximating the Frobenius-Perron operator would appear 
as the deterministic map.   For this reason, we define the limit of $T$ matrices as
 $K\rightarrow \infty$ to be a singular limit of the Frobenius Perron operators, and associated the associated transformations.

There is an interesting correspondence between the spectrum of eigenvalues of
the two matrices $D$ and $T$.
Since $T$ is stochastic its largest eigenvalue is equal to unity.
Moreover, it is the only eigenvalue with modulus one, 
which follows from the fact that
the kernel ${\cal P}(x,y)$ vanish only for $x-y=1/2$ (mod $1$),
and the two--step probability function is everywhere positive, 
\begin{equation}
\int {\cal P}(x,z){\cal P}(z,y) dy > 0,\mbox{ for }x,y\in X.
\end{equation}
See \cite{Lasota-Mackey},~Th.~5.7.4. 
A particularly useful consequence and simplification is that the eigenstate
corresponding to the largest eigenvalue 
of the matrix
represents the invariant density of the system, $\rho_*=P_f(\rho_*)$; this can
be easily found  numerically by diagonalizing $T$. 

All of the other eigenvalues are included inside the unit circle
and their moduli $|\lambda_i|$
characterize the decay rates.
It is worth emphasizing that
the spectra of both matrix representations of the 
FP-operator - by matrices  $D$ of size $(N+1)\times (N+1)$ used 
in \cite{OPSZ00,OZ01,OMZH03}
and the stochastic $T$ matrices of size $(N/2+1)\times (N/2+1)$, developed
here, 
coincide up to the additional  $N/2$ eigenvalues which are equal to zero -- see
the Appendix for details. 

For concreteness let us discuss an exemplary 1-D dynamical system,
a tent map: 
\begin{equation}
f(x) :=  \left\{  \begin{array}{cc}
                2x     \quad  &  {\rm if \quad } 0\le x\le 1/2  \\
                2(1-x) \quad &    {\rm if \quad } 1/2 \le x \le 1 \: . \\
                    \end{array} \right.
\label{tent}
\end{equation}
Simple integration allows us to obtain
analytic form of the transition matrix $T^{(N)}$
for the tent map (\ref{tent})
perturbed by additive noise
characterized by small values of $N$,
\begin{equation}
T^{(2)}=
\frac{1}{2}
\left[
\begin{array}{c c }
 1 & 1 \\
 1  & 1
\end{array}
\right] ,
\quad
T^{(4)} =
\frac{1}{24}
\left[
\begin{array}{c  c c}
 11 & 3   & 11 \\
 6   & 6  & 6  \\
 7   & 15  & 7
\end{array}
\right] , 
\quad
T^{(6)} =
\frac{1}{320}
\left[
\begin{array}{c c c c}
 145  & 25  &  25  & 145 \\
 69   & 45  &  45  & 69  \\
 51   & 75  &  75  & 51  \\
 55   & 175 & 175  & 55
\end{array}
\right] .
\label{TTT2}
\end{equation}

In the simplest case $N=2$ the transition 
matrix is bistochastic, but it is not so for larger $N$.
However, for this system, the matrix
$T^{(N)}$ is of rank one for arbitrary value of the noise parameter $N$.
The spectrum of $T$ contains one eigenvalue equal to unity
and all others equal to zero. This implies that every 
initial density is projected onto
an invariant density already
 after the first iteration of the map.
This is not the case for other dynamical systems $f$
including the logistic map $f_r(x)=rx(1-x)$,
for which the spectrum contains
several resonances - eigenvalues of moduli smaller than one,
which describe the decaying modes of the system 
\cite{OPSZ00}.

\section{Approximation by Basis Markov Maps}

While not all maps and noise profiles allow for the map to be basis Markov, in
this section we will show that a non basis Markov map may be weakly well
approximated by basis Markov maps.  In  this sense, the finite approximations
offered by basis Markov maps can be thought of as a good description of the
general behavior, since the invariant measures of the finite approximations
due to the basis Markov maps have weak-* limits to the invariant measures of
the general maps. 

Considering the transition probabilities in Eqs.~(\ref{prob1})-(\ref{prob2}),
we now write $\mathcal P_N(x,y)$ to denote the subindex $N$ to describe the
finite number of terms sufficient to describe the probability in assumption
(\ref{prob2})c.
We require the following assumptions about the transition probabilities
$\mathcal P_N(\cdot, \cdot)$:
\begin{enumerate}
\item $\mathcal P_N(\cdot,\cdot)$ is measurable as a function of two variables;
\item For every $x$ we have, $\int_0^1 \mathcal P_N(x,y)dy =1$.
\item For every $y\in X$ we have $$\int_X \mathcal P_N(x,y)dx =1.$$
\item Let $B(x,r)=\{y:|x-y|< r\}$ and,
\begin{equation} p_N(x,r)=\int_{X\setminus B(x,r)}\mathcal P_N(x,y) dy.
\end{equation}
Then, for any $r>0$,
 $$p_N(r)=\sup_{x\in X} p_N(x,r)\to 0,\ \ \ \text {  as  }\ \ \  N\to
+\infty.$$
\end{enumerate}

Assumptions 1-3 are typical for probability measures, while assumption 4 is
also rather mild, and it is easy to check that all four assumptions are
satisfied by the cosine noise Eq.~(\ref{trignoise}).

Under these assumptions, the following is true:

\medskip
{\bf Proposition:} For any $\rho\in L^1(X)$ we have
\begin{equation} \int_{X}\rho(x)\mathcal P_N(x,y) dx \to \rho(y), \mbox{ as }
N\rightarrow \infty
\end{equation}
 in $L^1$(X).
\medskip

{\it Proof:} Let us assume that $\rho$ is uniformly continuous and let us fix
an $\varepsilon>0$.
We can find an $r>0$ such that $|\rho(x)-\rho(y)|<\varepsilon$ whenever
$|x-y|<r$. We have
\begin{eqnarray}
& \int_X |\rho(y)-\int_X\rho(x)\mathcal P_N(x,y)dx|dy  =\nonumber\\
 (\text{assumption (3)})\  = 
& \int_X |\int_X\rho(y)\mathcal P_N(x,y)dx-\int_X\rho(x)\mathcal P_N(x,y)dx|dy 
\nonumber \\
& \le  \int_X\int_X |\rho(y)-\rho(x)|\mathcal P_N(x,y)dxdy \nonumber\\
&  = \int \int_{\{(x,y):|x-y|<r\}} |\rho(y)-\rho(x)|\mathcal P_N(x,y)dxdy
\nonumber \\
&  +\int \int_{\{(x,y):|x-y|\ge r\}} |\rho(y)-\rho(x)|\mathcal
P_N(x,y)dxdy\nonumber\\
 \nonumber\\
& \le  \varepsilon\cdot 1 + 2\cdot(\max_X |\rho|)\cdot p_N(r).
\end{eqnarray}
The last estimate can be made arbitrarily small in view of assumption  (4). 

Once the convergence is proven for uniformly continuous functions,  the proof
of convergence for general
$L^1(X)$ functions is standard: 

Since the natural norm of the operator $P_{\mathcal
P_N}(\rho)(y)=\int_X\rho(x)\mathcal P_N(x,y)dx$ in $L^1(X)$
is 1,  we have
\begin{eqnarray}
&\|\rho-P_{\mathcal P_N}(\rho)\|  \le
\|\rho-\rho_c\|+\|\rho_c-P_{\mathcal P_N}(\rho_c)\|+\|P_{\mathcal
P_N}(\rho_c)-P_{\mathcal P_N}(\rho)\| \nonumber \\
&\le 
\|\rho-\rho_c\|+\|\rho_c-P_{\mathcal P_N}(\rho_c)\|+\|P_{\mathcal
P_N}\|\|\rho_c-\rho\| \nonumber \\
&\le 2\|\rho-\rho_c\|+\|\rho_c-P_{\mathcal P_N}(\rho_c)\|,
\end{eqnarray}
where $\rho_c$ is a uniformly continuous approximation of $\rho$. $\Box$
\medskip

We will say that the transformation $f:[0,1]\to[0,1]$ preserves continuity
iff for any continuous function $\rho$ the composition $\rho\circ f$ is also continuous.
Obviously any continuous transformation $f$ preserves continuity. There are discontinuous
maps which also preserve continuity, e.g., $f(x)=kx -{\rm Int}(kx)$, for integer $k\ge 2$.
We now confirm the following result, stated in the language of our current
problem,

\medskip

{\bf Theorem: 1}. Let the transformation $f$ be continuity preserving. 
Under the assumptions (1), (2), and (4), it follows that
 if $\mu_N$ is an invariant measure of the stochastic perturbation of
transformation $f$ defined by
the transition probability $\mathcal P_N$, then every weak-$*$ limit point of
the set $\{\mu_N: N\ge 1\}$
is an $f$-invariant measure.
\medskip 

{\it Proof:}
This theorem can be proved following the ideas from, R.Z. Khasminskii,
\cite{Khasminskii}. Let us assume that $\mu_N\to \mu$ weakly as $N\to+\infty$.
We want to show that $\mu$ is $f$-invariant. To this end it is enough to show that
$$\int_X \rho d\mu = \int_X \rho(f) d\mu,$$
for any continuous function $\rho$. The stochastic perturbation of $f$ defined by
transition probability $\mathcal P_N$ acts on continuos functions as a compositions
with $f$ followed by  application of the operator ${P^*}_{\mathcal P_N}$, 
defined as follows
$$( {P^*}_{\mathcal P_N}\rho)(x)=\int_X \rho(y)\mathcal P_N(x,y) dy.$$
This operator is conjugated to the operator  ${P}_{\mathcal P_N}$ defined in
 the proof of the previous theorem. ${P^*}_{\mathcal P_N}$ acts on functions, while
${P}_{\mathcal P_N}$ acts on functions understood as densities.
Thus,
$$ \int_X \rho d\mu_N = \int P^*_{\mathcal P_N}(\rho(f)) d\mu_N,$$
for any continuous function $\rho$.
Using assumption (4) we obtain 
\begin{eqnarray*}
&|\rho(x)-(P^*_{\mathcal P_N}\rho)(x)|\\
&=|\int_X \rho(x)\mathcal P_N(x,y) dy-\int_X\rho(y)\mathcal P_N(x,y) dy|\\
&\le \int_{\{y:|y-x|<r\}}|\rho(x)-\rho(y)|\mathcal P_N(x,y) dy+
\int_{\{y:|y-x|\ge r\}}|\rho(x)-\rho(y)|\mathcal P_N(x,y) dy\\
&\le \varepsilon+ 2\cdot(\max_X |\rho|)\cdot p_N(r),
\end{eqnarray*}
where $\varepsilon$ and $r$ are as in the proof of the previous theorem. This shows that
$P^*_{\mathcal P_N}(\rho)$ converges uniformly to $\rho$ for any continuous function $\rho$ as
$N\to +\infty$.

 We have
\begin{eqnarray*}
&|\int_X \rho d\mu - \int_X \rho(f) d\mu| \le |\int_X \rho d\mu - \int_X \rho d\mu_N|\\
& + |\int_X \rho d\mu_N - \int P^*_{\mathcal P_N}(\rho(f)) d\mu_N|+
|\int P^*_{\mathcal P_N}(\rho(f)) d\mu_N-\int_X \rho(f) d\mu_N|\\
&+|\int_X \rho(f) d\mu_N-\int_X \rho(f) d\mu|
\end{eqnarray*}
The first and the last differences converge to 0 since $\mu_N$ converge $*$-weakly  to 
$\mu$. The second difference is 0 by the definition of $\mu_N$. The third difference
converges to 0 by the uniform convergence established just before.
This proves Theorem 1. $\Box$
\medskip

In this way we have established  a relation between a 
sequence of noisy systems $f_N$
and the deterministic dynamical system $f$.
A stochastic system (\ref{addit})
with the noise profile (\ref{trignoise}) for 
a fixed noise parameter $N$ is
described by a stochastic matrix $T^{(N)}$ of size $K=N/2+1$
and acts in the Hilbert space ${\cal H}_K$.

We have shown that the sequence of stochastic matrices $T^{(N)}$,
{\sl corresponds} to the dynamical system $f$,
in a sense that the sequence $\mu_N$
of the invariant measures of $T^{(N)}$
converge weakly to the  $f$-invariant measure $\mu$
in the 
deterministic limit $N\to \infty$.
Furthermore, for any initial density $\rho$ the sequence of
vectors $\rho'_N$ transformed by $f_N$ converges weakly to the 
density transformed by the Frobenious--Perron operator
associated with $f$.
Observe that the above property holds not only for one-dimensional systems, but also  
dynamical system $f$ in higher dimensional measure spaces.

\section{Concluding Remarks}

In this work we have introduced the concept of basis--Markov
stochastic systems, for which the associated Frobenius--Perron 
operator is finite. This property resembles the class of deterministic
systems with a Markov partition. However, the Markov partition
is characteristic to a very special class of deterministic systems,
while the basis--Markov property is related to the kind of 
stochastic perturbation. It holds for any deterministic system $f$,
subjected to an additive noise with a profile satisfying 
the separability condition  (\ref{expand}).
In this way such a random dynamical system
can be described by a stochastic transition matrix of a finite size $K$,
which diverges in the deterministic limit.

We have shown an intimate relationship between the sequence of 
stochastic matrices which act in the space of $K$--point probability
distributions  and the FP operator $P_f$ of the deterministic system,
which acts in the infinite dimensional space: 
In the deterministic limit $K\to \infty$ 
the invariant densities of stochastic matrices 
converge in a weak sense to the invariant
measure of the deterministic system $f$. 
Thus constructing the transition matrices $T$
and decreasing the noise strength 
(and increasing the dimensionality $K$)
one may construct arbitrary approximations
of the FP operator $P_f$. 

Note that the described method of
finding an approximate invariant density of a deterministic system
by applying a weak noise is not restricted to one dimensional
systems. On the contrary, the entire construction
can be directly applied for a general case of multi--dimensional
dynamical systems.
In particular, the definition (\ref{expand})c of separable noise profiles
works for the case of an $L$--dimensional systems,
provided the variables $x$ and $y$ represent vectors with $L$ components each.

If the dynamical system acts on the $L$--torus for example, $M=[0,1]^L$,  
one can take the Cartesian product of the cosine noise
(\ref{trignoise}) setting
\begin{equation}
{\cal P}_N(\xi_1,\dots\xi_L) =C_N^L \cos^N(\pi\xi_1)
\cos^N(\pi\xi_2) \cdot \cdot
\cos^N(\pi\xi_L) \ ,
\end{equation}
where $\xi_k=x_k-y_k$ and $k=1,\dots, L$.
This form of the additive noise was used in \cite{OMZH03}
to analyze a $2$-dimensional system 
(a variant of the baker map), and to compare the 
spectral properties of the FP operator associated
with the classical stochastic system
with properties of the propagator of the corresponding
quantum evolution. In such a case the deterministic
limit of the classical noisy system, $K\to \infty$
is related to the classical limit, $\hbar \to 0$,
of the corresponding quantum dynamics.

Note that for basis--Markov stochastic systems, the transition matrices $T$
{\sl exactly} describe  the action of the dynamical system with  additive noise on densities.
Thus our construction  differs from an approach applied 
in \cite{fishman,weber2,Nonnenm}, were a finite dimensional
description of the density dynamics  of a deterministic system
was achieved by truncation of an infinite transition operator  $P_f$
to the finite dimension $K$. 
The effect of such a truncation may also be regarded as
a kind of noise depending on the matrix size $K$ and 
the base, in which $P_f$ is represented. 
On the other hand, in our case a suitable choice of the noise profile added to
the deterministic system distinguishes a relevant basis, in which
the FP operator of the perturbed system is finite.

\section{\large    Acknowledgments}\label{ack}

E. B. was supported by the National Science Foundation under grants
DMS-0404778 while K. {\.Z}. acknowledges a partial support by the grant number
1\, P03B\, 042\, 26 of Polish Ministry of Science and Information Technology.

\section{Appendix: isospectral matrices}

In this appendix we show that the matrix $D$
defined by  Eq.~(\ref{drm}) and used in \cite{OPSZ00,OZ01,OMZH03}
to represent the Frobenius-Perron operator
and the stochastic transition matrix $T$
share the same non-zero part of the spectrum.
We make use of a following algebraic result,
\medskip

{\bf Lemma.} {\sl Let $A$ be a square 
matrix of size $N \times N$ and $\vec s$ a vector of length $N$
containing only non-zero entries. Then the matrix}
\begin{equation}
B_{jk} \ \equiv\  A_{jk} \, \frac{s_j}{s_k}, 
\label{AB}
\end{equation}
 {\sl 
has the same spectrum as $A$}.
\medskip

(there is no summation over repeating indices).
\medskip

{\it Proof:}  To study equation  
det$(B-\lambda {\mathbbm 1})=0$
we start analyzing an exemplary term $P^B$ of the determinant.
It consists of a product of $N$ elements $B_{i,\sigma(j)}$,
where $\sigma(i)$ stands for a certain permutation of the indeces.
The product of $N$ factors of the type ${s_i}/s_{\sigma(i)}$
is equal to unity, so that  
\begin{equation}
P^B_{\sigma}=\prod_i B_{i,\sigma(i)}=\prod_i B_{i,\sigma(i)}\,
\frac{s_1 s_2 \cdots s_N} {s_1 s_2 \cdots s_N}
 =\prod_i A_{i,\sigma(i)} \ .
\label{psigma}
\end{equation}
Thus every term contributing to the free coefficient
of the characteristic equation
 will  be the same, $P^B_{\sigma}=P^A_{\sigma}$,
hence these coefficients for both matrices $A$ and $B$
are equal. Since the diagonal elements of both matrices
coincide, $B_{jj}=A_{jj}$, all terms
forming the coefficients standing
by an arbitrary power of $\lambda$
are the same for both matrices.
Therefore characteristic equations
for both matrices are equal and so are their spectra. $\square$.

Treating all non-zero  elements of the vector $ \tau_k, 
k=1,\dots,K$ as vector $\vec s$
we may apply the lemma 
to equation (\ref{norm22})
and obtain equivalence of the spectrum of $T$
and the non-zero part of the spectrum of $D$.
Since integrals (\ref{brm})
vanish for odd values of $m$,
every second column of $D$ is equal to zero,
and the remaining  $N/2$ eigenvalues
of $D$ are equal zero.


\begin{thebibliography}{}
\bibitem{Bowen75} R. Bowen, {\sl Equilibrium States and the Ergodic
Theory of Anosov Diffeomorphisms}, Springer-Verlag, Berlin 1975.

\bibitem{BG}
A.~Boyarsky and P.~G\'{o}ra, \emph{Laws of chaos: invariant measures and
  dynamical systems in one dimension}, Birkhauser, Boston, 1997.

\bibitem{bollt33}
E. Bollt, J. Skufca, ``Markov Partitions," Encyclopedia of Nonlinear Science,
invited short, Editor, Alwyn Scott, Fitzroy Dearborn Publishers, to appear
(2003).   Available at http://www.clarkson.edu/$\sim$bolltem

\bibitem{Kitchens}
B.P.~Kitchens, {\it Symbolic Dynamics, One-sided, Two-sided and Countable
State Markov Shifts,}  Springer (New York, 1998).

\bibitem{Fischer}
R. Fischer, Sofic Systems and Graphs, Monatsh. Math. 80 (1975), 179-186.

\bibitem{Lasota-Mackey}
A. Lasota, M.C. Mackey, {\it Chaos, fractals, and noise, 2nd Ed.}
(Springer-Verlag, New York, 1994).

\bibitem{LArnold}
 L. Arnold, Random  Dynamical Systems, Springer-Verlag, New York, 1998.

\bibitem{Li}
T.-Y. Li, ``Finite Approximation for the Frobenius-Perron Operator. A Solution
to Ulam's Conjecture," J.~Approx.Th. {\bf 17}, 177-186 (1976).

\bibitem{Ulam}
S. Ulam, {\it Problems in Modern Mathematics,} Interscience Publishers, (New
York, NY, 1960).

 \bibitem{Robinson}
 C. Robinson, {\it Dynamical Systems, Stability, Symbolic Dynamics, and Chaos,
2$^{nd}$}, CRC Press, (1999).
 
 \bibitem{GuckHolmes}
 J. Guckenheimer, P. Holmes, {\it Nonlinear Oscillations, Dynamical Systems,
and Bifurcations of Vector Fields,} Springer (1991).

 \bibitem{OPSZ00}
A. Ostruszka, P. Pako{\'n}ski, W. S{\l}omczy{\'n}ski, and K. {\.Z}yczkowski, 
``Dynamical entropy for systems with stochastic perturbations",
 Phys. Rev. E, {\bf 62} 2 2018-2029 (2000). 

\bibitem{OZ01}   
A. Ostruszka, K. {\.Z}yczkowski,
 ``Spectrum of Frobenius-Perron operator for systems with stochastic
perturbation,"
 Phys. Lett. A {\bf 289} 306-312 (2001).

\bibitem{Tu00}R. R. Tucci,
 Entanglement of Formation and Conditional Information Transmission, 
preprint quant-ph/0010041 (2000).

\bibitem{OMZH03}
 A. Ostruszka, Ch. Manderfeld, K. {\.Z}yczkowski, and F. Haake,
"Quantization of Classical Maps with tunable Ruelle-Pollicott Resonances",
{\sl Phys. Rev. } {\bf E 68},  056201  (2003)

 \bibitem{Khasminskii} R.Z. Khasminskii, 
"Ergidicheskie svoistva vozvratnykh diffuzionnykh protsessov...",
Teoria Veroyat. i yeye Prim. 5, No. 2, (1960), 196--214.

\bibitem{fishman} M.~Khodas, S. Fishman, 
  Relaxation and Diffusion for the Kicked Rotor
{\sl Phys.~Rev.~Lett.} {\bf 84}  2837 (2000).

\bibitem{weber2} J.~Weber, F.~Haake, P.~A.~Braun, C.~Manderfeld and
  P.~\v{S}eba, 
Resonances of the Frobenius-Perron Operator for a Hamiltonian Map with a Mixed
Phase Space
{\sl J.~Phys.~A} {\bf 34} 7195 (2001).  

 \bibitem{Nonnenm} S.~Nonnenmacher, 
 Spectral properties of noisy classical and quantum propagators
{\sl Nonlinearity} {\bf  16}, 1685-1713 (2003).


\end{thebibliography}
\end{document}